\documentclass[aps,prl,twocolumn]{revtex4}
\usepackage{graphicx}
\usepackage{bm}
\usepackage{amsmath}
\usepackage[colorlinks=true,allcolors=blue]{hyperref}

\newcommand{\ket}[1]{|{#1}\rangle}

\newcommand{\kb}[2]{|#1\rangle\langle#2|}
\newcommand{\hH}{\hat{\mathcal{H}}}
\newcommand{\hrho}{\hat{\rho}}
\newcommand{\hU}{\hat{U}}
\newcommand{\sz}{\hat{\sigma}_{z}}
\newcommand{\sx}{\hat{\sigma}_{x}}

\newcommand{\sma}{\hat{\sigma}_{+}}
\newcommand{\sme}{\hat{\sigma}_{-}}

\begin{document}

\title{Pulse engineering for population control under dephasing and dissipation}

\author{I. Medina}

\address{Centro de Ci\^encias Naturais e Humanas, Universidade Federal do ABC,
09210-170, Santo Andr\'e, S\~ao Paulo, Brazil.}

\author{F. L. Semi\~ao}

\address{Centro de Ci\^encias Naturais e Humanas, Universidade Federal do ABC,
09210-170, Santo Andr\'e, S\~ao Paulo, Brazil.}

\begin{abstract}
\noindent We apply reverse-engineering to find electromagnetic pulses that allow for the control of populations in quantum systems under dephasing and thermal noises. In particular, we discuss two-level systems given their importance in the description of several molecular systems as well as quantum computing. Such an investigation naturally finds applications in a multitude of physical situations involving the control of quantum systems. We present an analytical description of the pulse which solves a constrained dynamics where the initial and final populations are fixed \textit{a priori}. This constrained dynamics is sometimes impossible and we precisely spot the conditions for that. One of our main results is the presentation of analytical conditions for the establishment of steady states for finite coherence in the presence of noise. This might naturally find applications in quantum memories.
\end{abstract}
\maketitle
\textit{Introduction} - The development of new techniques to control quantum systems is of fundamental importance 
to quantum technologies. 
Typical control protocols usually involve the interaction of the system of interest with external electromagnetic radiation \cite{key-9,key-10,key-11,key-12}. 
In particular, given that ultrashort laser pulses are now an experimental reality, femtosecond pulses are extensively used to control molecular dynamics \cite{key-13,key-14}. In this scenario, as surprising as it can sound, two-level systems often provide a powerful testbed to understand complicated molecular processes \cite{key-6,key-7,key-8}. For instance, the coupling of protein motion to electron transfer
in a photosynthetic reaction center is usually thought of as a legitimate spin-boson system \cite{sp}. Quantum control has been applied to these systems, for instance, to engineer specific pulses to control the populations. In \cite{key-16,key-17} it is discussed the pulse envelop form able to drive the state populations to an specific user-defined value. 

This inverse engineering approach, i.e., the design of controlled pulses or Hamiltonian parameters to satisfy dynamical constraints is by itself an interesting topic \cite{old,new}. When aiming at applications in molecular or condensed matter systems, it is necessary to go beyond closed systems and pure states. Our goal is to bridge this gap and present robust pulse control protocols that take into account the presence of the noise environments and the realistic use of general mixed states as initial states. Without these developments, it would be impossible, for instance, to included initial thermal equilibrium states into the problem. As it is going to be discussed in this work, noise drastically change the scenario found in \cite{key-16,key-17}. We come up with a broadly applicable platform for the control of populations under dephasing and thermal noise which allows us to establish clear bounds for the achievable populations in terms of the environmental features. We present several examples with a thoroughly discussion of about validity of our protocol. i.e., our analytical formula for the pulse. We finish with the presentation of  conditions for the achievement of steady states with finite values of quantum coherence in the presence of the environment.

\begin{figure}
\includegraphics[scale=0.3]{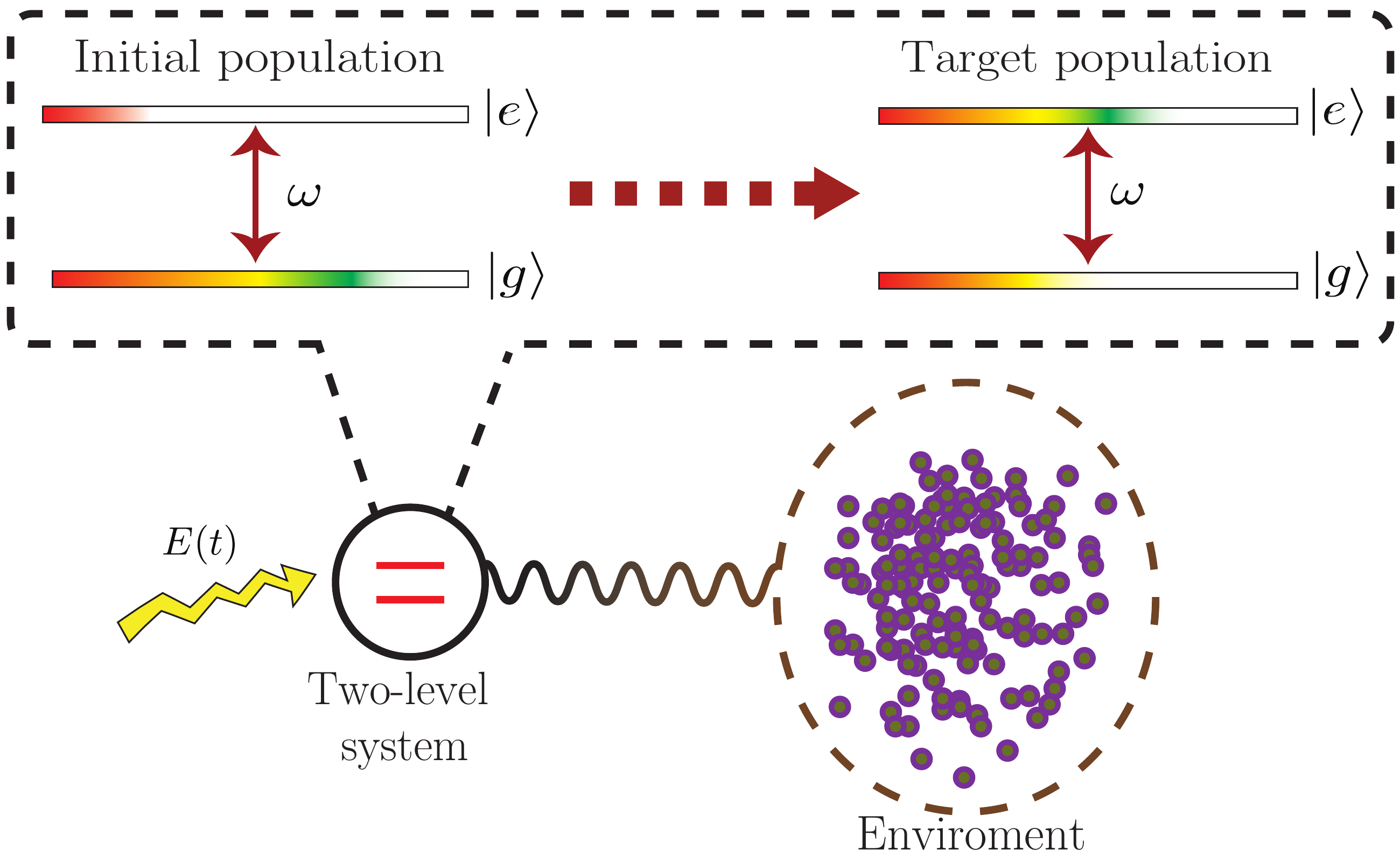}

\caption{(Color online) A two-level system representing, for instance, a coupled donor-acceptor system, interacts simultaneously with the environment and an externally controlled electromagnetic pulse $E(t)$. The purpose of the pulse is to induced a pre-determined change of population. The populations are indicated by the filling of the bars corresponding to the ground
$\protect\ket g$ and excited $\protect\ket e$ states. \label{fig:esquema} }
\end{figure}

\textit{Reverse-engineering method} - The situation we have in mind is depicted in Fig. \ref{fig:esquema}. In the dipole approximation, the system-pulse Hamiltonian reads $(\hbar=1)$
\begin{equation}
\hH(t)=-\frac{\omega}{2}\sz-\mu E(t)\sx,\label{h1}
\end{equation}
where $\sx$ and $\sz$ are usual Pauli operators, $\omega$ is the system transition
frequency, $E(t)$ is the pulse electric field, and
$\mu$ is the projection of the system electric dipole operator on the field polarization
direction. Atomic units are used throughout the text unless otherwise specified. In the case of an electromagnetic pulse with carrier frequency $\omega_{p}$,
the field can be written as
\begin{equation}
E(t)=\varepsilon(t)e^{-i\omega_{p}t}+\varepsilon^{*}(t)e^{i\omega_{p}t},\label{eq:field}
\end{equation}
where $\varepsilon(t)$ is a complex function which contains the amplitude
and envelope of the pulse. For convenience, we will treat the problem
in the interaction picture with respect to the unitary operation
$\hU=\exp[i\frac{\omega}{2}\sz(t-t_{0})]$.

The environment causes thermal noise and dephasing with rates $\Gamma$ 
and $\gamma$, respectively. In this scenario, the density matrix $\hrho$ obeys \cite{key-18}
\begin{equation}
\frac{\partial\hrho}{\partial t}=-i[\hat{V},\hrho]+\frac{\gamma}{2}D_{deph}[\hrho]+\Gamma D_{therm}[\hrho],\label{eq:master eq}
\end{equation}
where $\hat{V}=-\mu[\varepsilon(t)\sma e^{-i\Delta t-i\omega t_{0}}+h.c]$
is the Hamiltonian in the interaction picture and under the rotating
wave approximation (RWA),  $\Delta=\omega_{p}-\omega$, 
$D_{deph}[\hrho]=\sz\hrho\sz-\hrho$, and $D_{therm}[\hrho]=\bar{n}(2\sma\hrho\sme-\{\sme\sma,\hrho\})+(\bar{n+1})(2\sme\hrho\sma-\{\sma\sme,\hrho\})$,
where $\sma=(\sme)^{\dagger}=\kb eg$ and $\bar{n}$ is the average
number thermal phonons. It follows from this master equation that
\begin{align}
\dot{\rho}_{gg} & =2\mu\text{Im}[\rho_{eg}\varepsilon(t)e^{-i(\Delta t+\omega t_{0})}]+2\Gamma[(\bar{n}+1)\rho_{ee}-\bar{n}\rho_{gg}],\label{eq:dif1}\\
\dot{\rho}_{eg} & =-\tilde{\Gamma}\rho_{eg}+i\mu[2\rho_{gg}-1]\varepsilon(t)e^{-i(\Delta t+\omega t_{0})},\label{eq:ref4}
\end{align}
where we defined $\rho_{ij}=\langle i|\rho|j\rangle$ with $i,j=g,e$, and $\tilde{\Gamma}=[\gamma+(2\bar{n}+1)\Gamma)]$
as the total decoherence rate. The reverse-engineering method consists
in obtaining the field $E(t)$ from this set of differential equations and under the desired constraints on the density matrix elements.
From Eq.~(\ref{eq:ref4}), we isolate $\varepsilon(t)e^{-i\omega_{p}t}$
and, upon using (\ref{eq:field}), we
obtain
\begin{equation}
E(t)=2\frac{\text{Im}\left[\left(\dot{\rho}_{ge}+\tilde{\Gamma}\rho_{ge}\right)e^{i\omega(t-t_{0})}\right]}{\mu(2\rho_{gg}-1)},\label{eq:fieldr}
\end{equation}
which is an expression for the field in terms of the density matrix
elements of the system.

Now we impose the desired constraints. Our goal is to find a pulse that promote the change from an initial ground state population $a_{i}=\langle g|\hrho(-\infty)|g\rangle$
to the ``user-defined''  final ground state population $a_{f}=\langle g|\hrho(\infty)|g\rangle$.
Keeping this in mind, we now constrain the density matrix element $\rho_{gg}$ to follow a prescribed time evolution $\rho_{gg}=f(t)$, where
\begin{equation}
f(t)=[1-g(t)]a_{i}+a_{f}g(t),\label{eq:fpop}
\end{equation}
with $g(t)$ chosen to be
$
g(t)=(1+e^{-\alpha t})^{-1},
$
and $\alpha$ a real and positive parameter that dictates the rate of the transition from $a_{i}$ to $a_{f}$  \cite{key-17}. For convenience, we rewrite the coherence $\rho_{ge}$ as
$\rho_{ge}=h(t)e^{i\phi(t)}$, with $h(t)\equiv|\rho_{ge}|$, requiring only that the fase is time independent, i.e., $\phi(t)=\phi_{0}$ with $\phi_{0}$  real. No requirements are imposed on $h(t)$ which is the state coherence. Now, by using  (\ref{eq:dif1}) and (\ref{eq:ref4}), one obtains
\begin{equation}
\dot{h}(t)=\frac{[2f(t)-1]}{2h(t)}[2\Gamma(1+\bar{n}-(2\bar{n}+1)f(t))-\dot{f}(t)]-\tilde{\Gamma}h(t),\label{eq:difh}
\end{equation}
which is a type of differential equation known as Bernoulli equation. 
Finally, using Eqs.~(\ref{eq:fieldr}),  (\ref{eq:fpop}) and (\ref{eq:difh}), one finds 
\begin{equation}
E(t)=\frac{1}{\mu h(t)}[\dot{f}(t)-2\Gamma(1+\bar{n}-(2\bar{n}+1)f(t))]\sin\theta(t),\label{eq:fieldf}
\end{equation}
with $\theta(t)\equiv\omega(t-t_{0})+\phi_{0}$. This equation constitutes the main result of this paper. It gives a recipe for the experimentalist to build a pulse to reach a target final ground state population under thermal and dephasing noise, i.e., under very realistic scenarios.

We now proceed to some quantitative simulations. The field Eq.~(\ref{eq:fieldf}) is substituted into Hamiltonian (\ref{h1}) and the equations of motion with dephasing and thermal noises are solved numerically with no RWA. For the initial state, we choose the broad class of diagonal states $\hrho(t_{0})=a_{i}\kb gg+(1-a_{i})\kb ee$ which include, for instance, the Gibbs state, in which the
populations are determined by the Boltzmann factor $\exp[-E_{n}/k_{B}T]$
where $T$ is the temperature, $k_{B}$ the Boltzmann constant
and $E_{n}$ are eigenenergies of the system. As an application, we chose $\omega=2\times10^{-2}$ a.u., $\alpha=10^{-2}$ a.u., $\mu$ = $6$ a.u. which corresponds to the physical situation of charge
migration in the molecule MePeNNA \cite{key-15}. Its ionization can trigger a ultrafast migration of charge which can be acurately
 modeled as a two level system \cite{key-15,key-16}. By including the environment, our approach represents a step forward in the direction of having more realistic models for this phenomenon.

\begin{figure}
\includegraphics[viewport=0bp 0bp 467bp 546bp,scale=0.4]{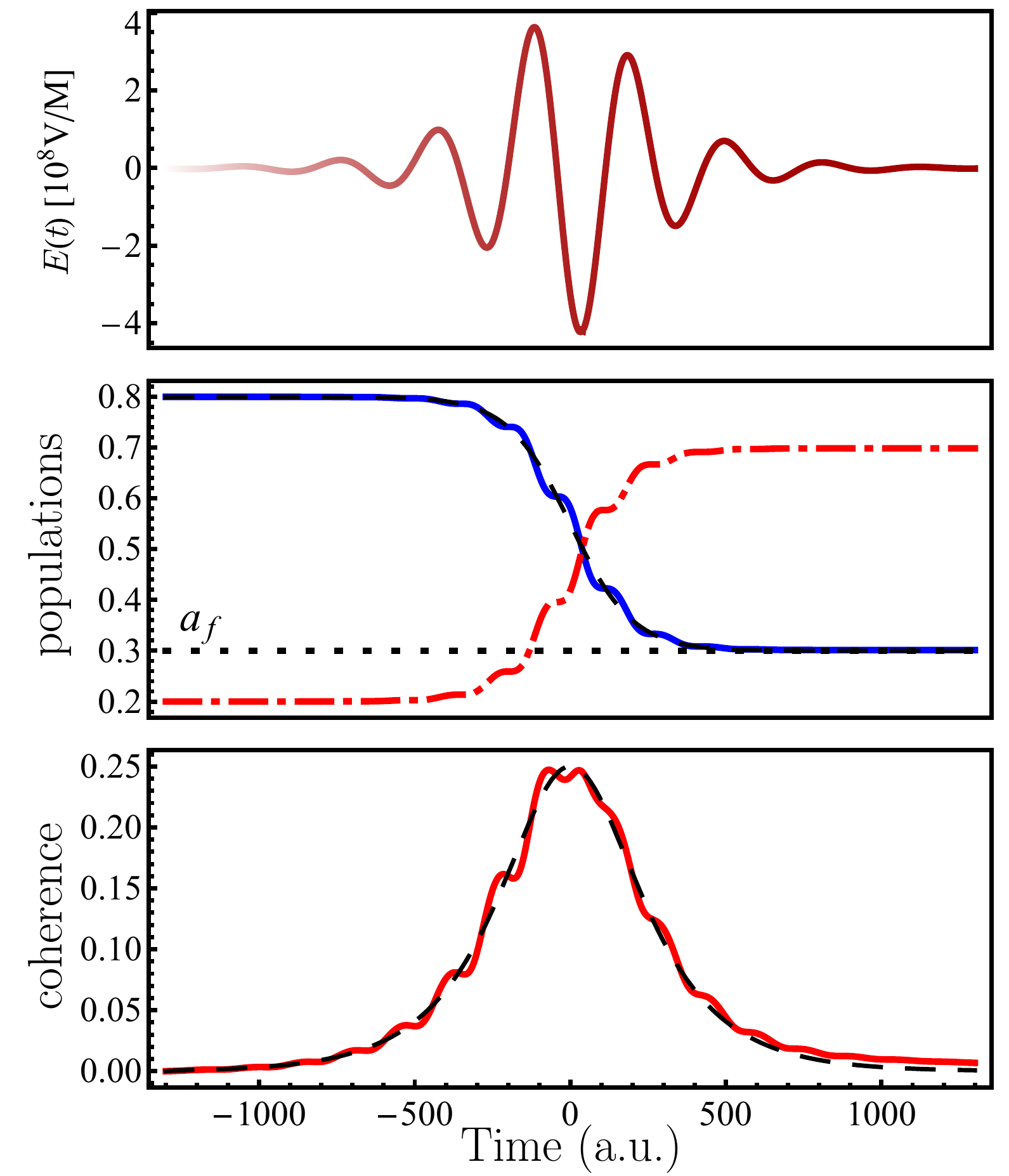}
\caption{(Color online)  \textbf{Top panel}: Laser pulse as Eq.~(\ref{eq:fieldf}). \textbf{Middle panel}:  Ground state population $\rho_{gg}(t)$
is shown in blue (solid), excited state population $\rho_{ee}(t)$ is shown
in red (dot-dashed) and the control function $f(t)$ is shown in black (dashed).\textbf{
Lower panel: }The absolute value of the coherence $|\rho_{eg}(t)|=|\rho_{ge}(t)|$ is shown in red (solid) and
the function $h(t)$ is shown in black (dashed). For all plots,  $a_{i}=0.8$,
$a_{f}=0.3$, $\Gamma=0$, and $\gamma=10^{-3} a.u.$.
\label{fig:1} }
\end{figure}

\textit{Pure dephasing} - In this case, we set $\Gamma=0$ so that only the coherence $\rho_{eg}(t)$ is directly
affected by the environment. However, one must not forget that at the same time the laser causes the populations to change, and the simultaneous action of dephasing and laser can cause non trivial dynamics as we now discuss. One illustrative example is depicted in Fig. \ref{fig:1}. The initial
population is given by $a_{i}=0.8$ and we set the target population
to $a_{f}=0.3$. The top panel shows the shape of the pulse obtained
with Eq.~(\ref{eq:fieldf}). The middle panel shows that despite the fact that Eq.~(\ref{eq:fieldf}) has been obtained in the RWA, the protocol works very well when this approximation is performed. In other words, the system closely follows the desired evolution given by $f(t)$. The superposed small amplitude oscillations introduce
only small deviations which are absolutely unimportant for the present task. The same occurs with
the coherence and the function $h(t)$ showed in the lower panel.
It is interesting to see that the interaction with the pulse generates
coherence which is eventually consumed by 
the dephasing which incoherently drives system to a final diagonal state. Later on in this article, we will show a very interesting case where the coherence is not completely degraded as $t\rightarrow\infty$.

\begin{figure}
\includegraphics[scale=0.38]{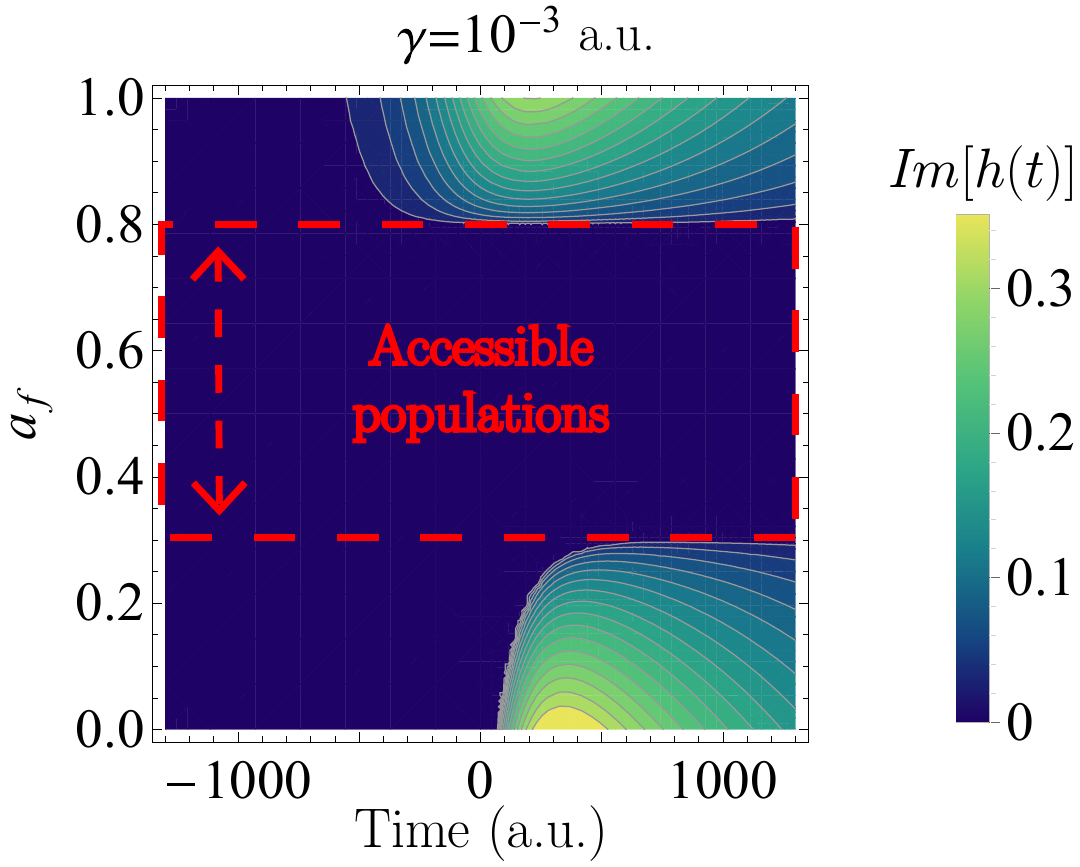}\includegraphics[scale=0.38]{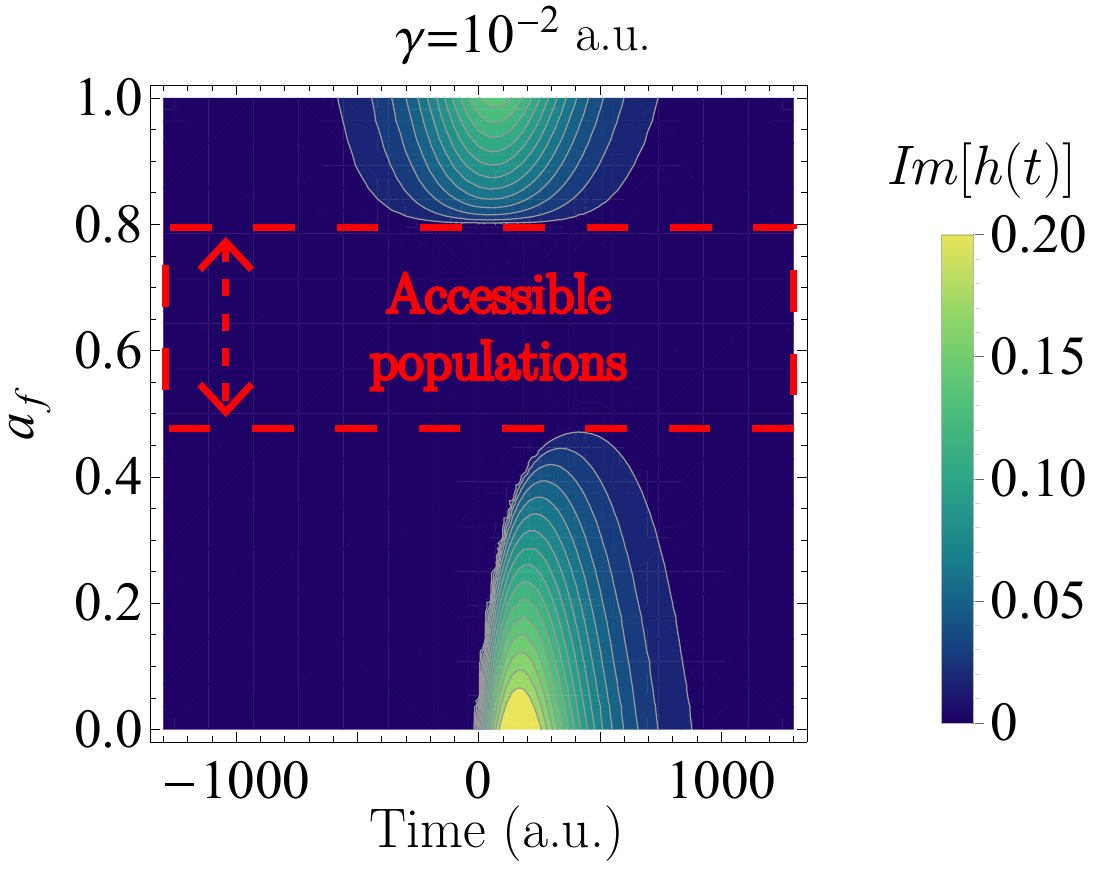}
\caption{(Color online)\textbf{ }Imaginary part of the function
$h(t)$ as function of time and the target final population $a_{f}$ for
two choices of $\gamma$. 
The frames in red (dashed) show the accessible region for $a_{f}$. \label{fig:2}}
\end{figure}

In Fig. \ref{fig:2}, we show contour plots of $h(t)$ as a function
of time and the target population $a_{f}$, for some choices of the dephasing rate
$\gamma$. We are fixing the initial ground state population as $a_{i}=0.8$.
We see that for some values of $a_{f}$, the function $h(t)$ assume complex values. Given that we have defined $h(t)$ as the absolute value
of the coherence, such regions correspond to nonphysical states or equivalently complex valued fields as seen from Eq.~(\ref{eq:fieldf}). Also, we can see that the higher
the value of $\gamma$ the smaller the region of accessible final populations.

%

\begin{figure}
\includegraphics[scale=0.4]{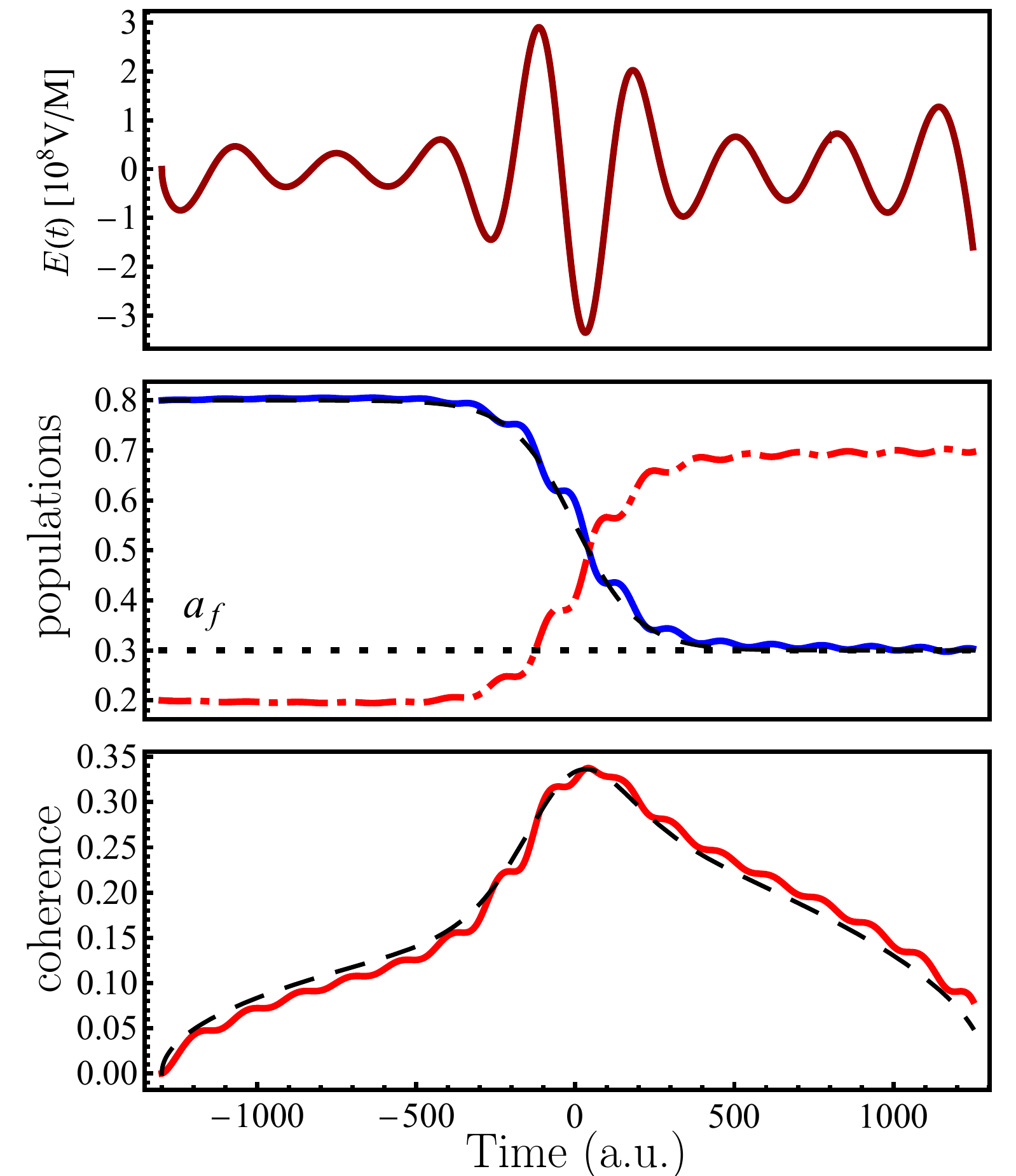}\caption{(Color online)  \textbf{Top
panel}: Laser pulse as Eq.~(\ref{eq:fieldf}). \textbf{Middle
panel}:  Ground state
population $\rho_{gg}(t)$ is shown in blue, excited state population
$\rho_{ee}(t)$ is shown in red (dot-dashed) and the control $f(t)$
is shown in black (dashed). \textbf{Lower panel: } The absolute value of the coherences $|\rho_{eg}(t)|=|\rho_{ge}(t)|$
are shown in red and the function $h(t)$ is shown in black (dashed). For all plots,  $a_{i}=0.8$,
$a_{f}=0.3$, $\Gamma=10^{-4}$ a.u., $\gamma=0$, $\bar{n}=0$.
\label{fig:5}}
\end{figure}

\textit{Thermal noise} - In this case, both the populations and the coherence are directly affected
by the environment. Also, the field acquires a dependence on $\Gamma$
and $\bar{n}$ as seen in Eq.~(\ref{eq:fieldf}). For the plots in Fig. \ref{fig:5}, we used $\Gamma=10^{-4}$ a.u., $\gamma=0$, $\bar{n}=0$, $a_{i}=0.8$
and $a_{f}=0.3$. The top panel shows that the field must quickly increase to counter the dissipation. After completing the
transition, the pulse strives to maintain the state with the desire population $a_f(t)$ in the presence of dissipation. This is possible only until
$h(t)$ reach, for a finite time, the value $0$. From this point on, the field as given by Eq.~(\ref{eq:fieldf}) diverges
and the system looses the desired population.
The middle and lower panels show that the evolution of the population
and coherence follow the path prescribed by the functions $f(t)$ and $h(t)$.

\begin{figure}
\includegraphics[scale=0.38]{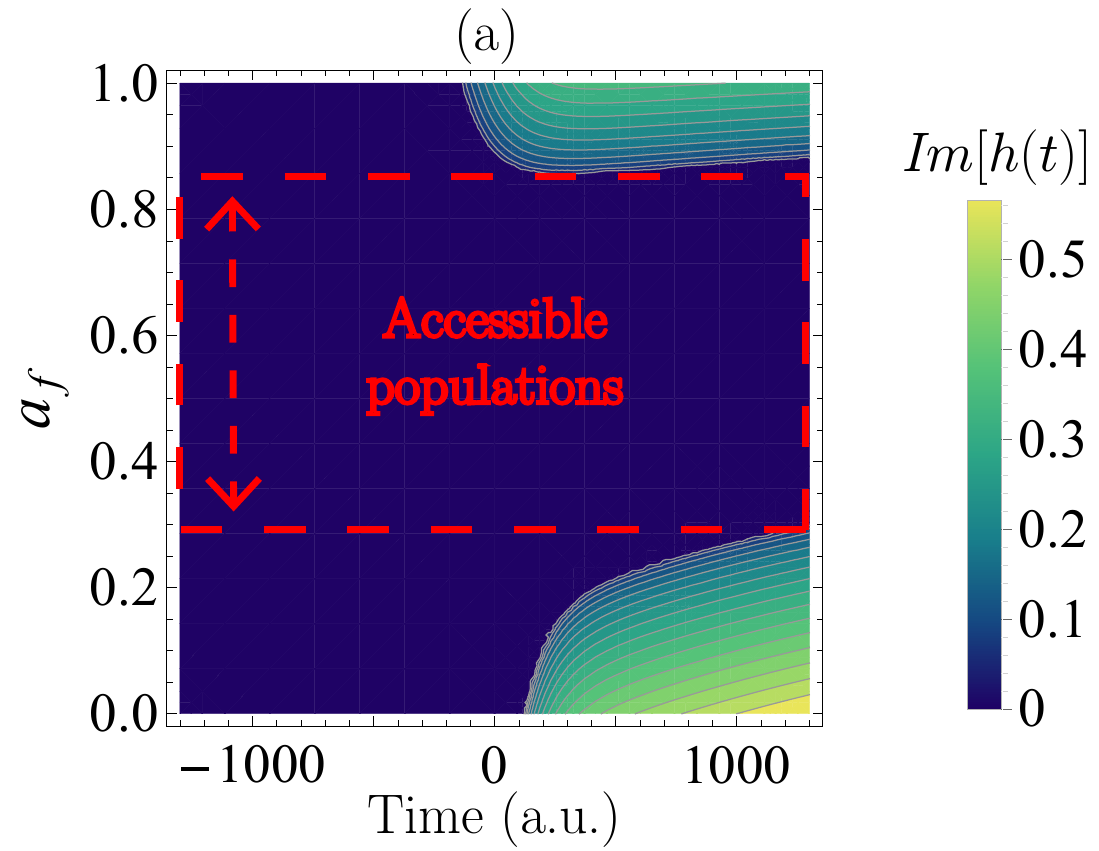}\includegraphics[scale=0.38]{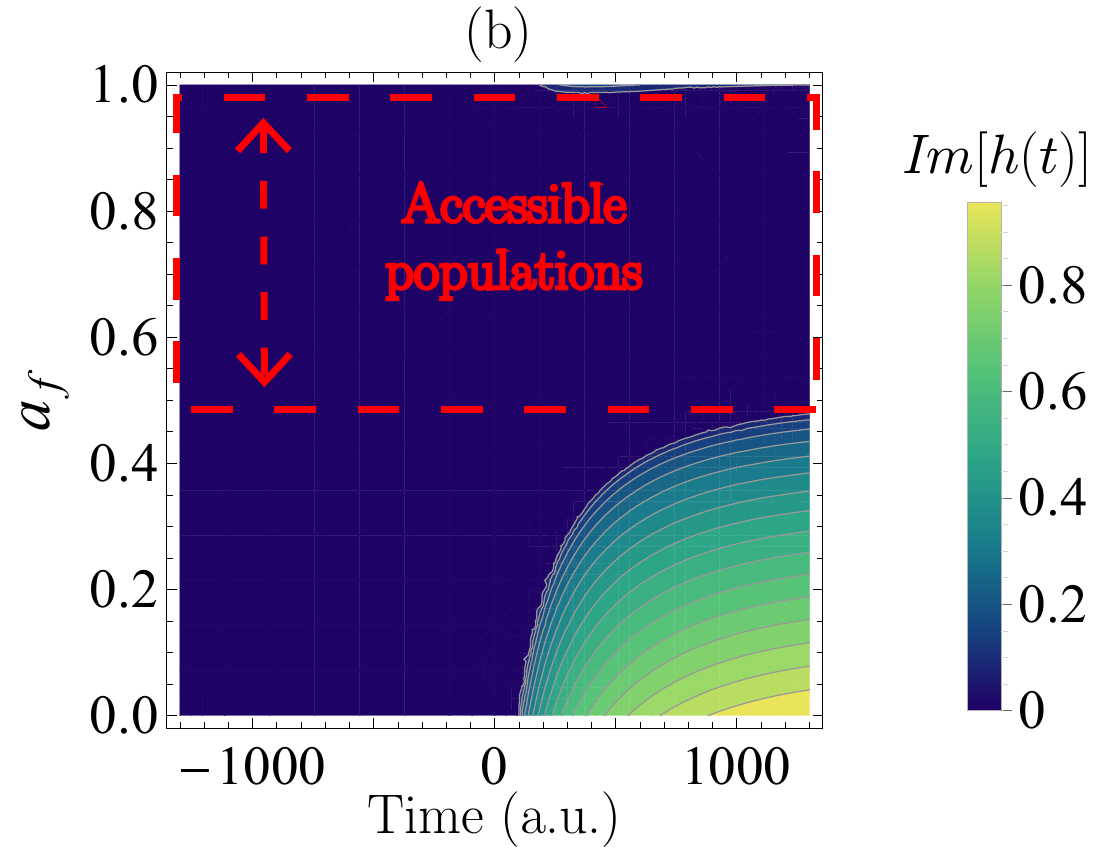}

\includegraphics[scale=0.38]{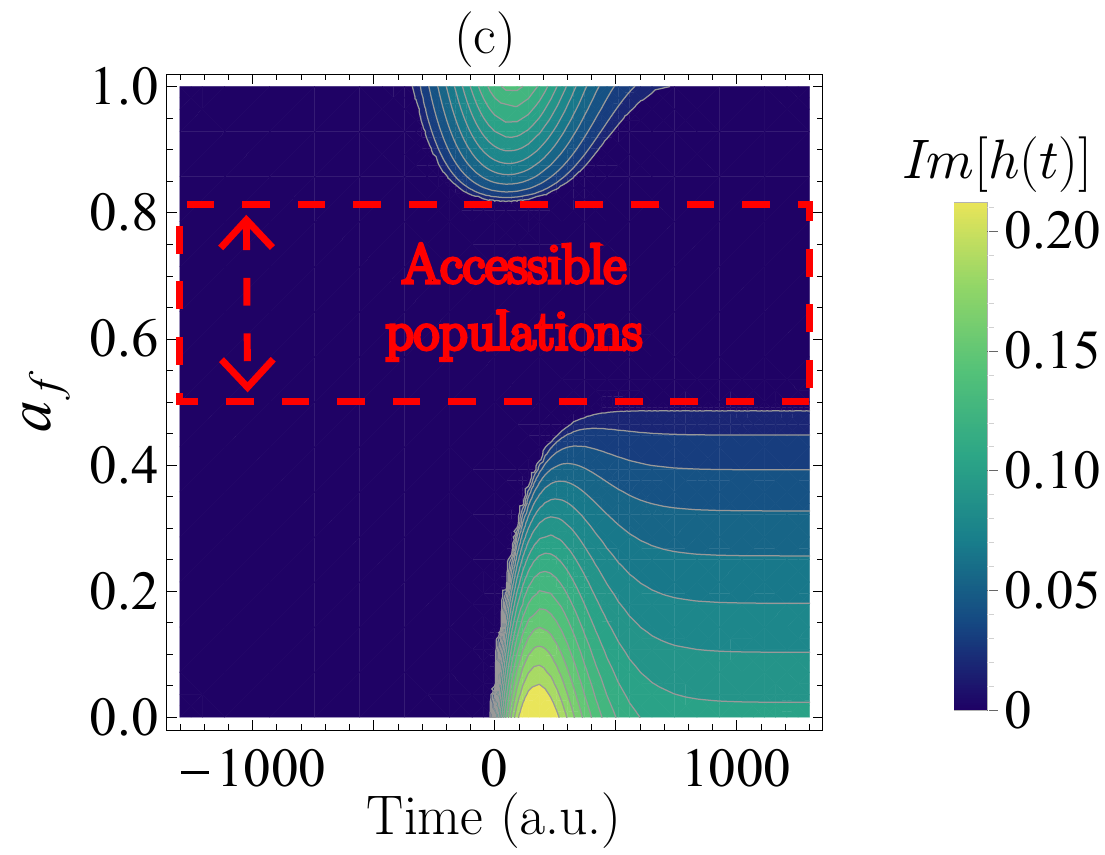}\includegraphics[scale=0.38]{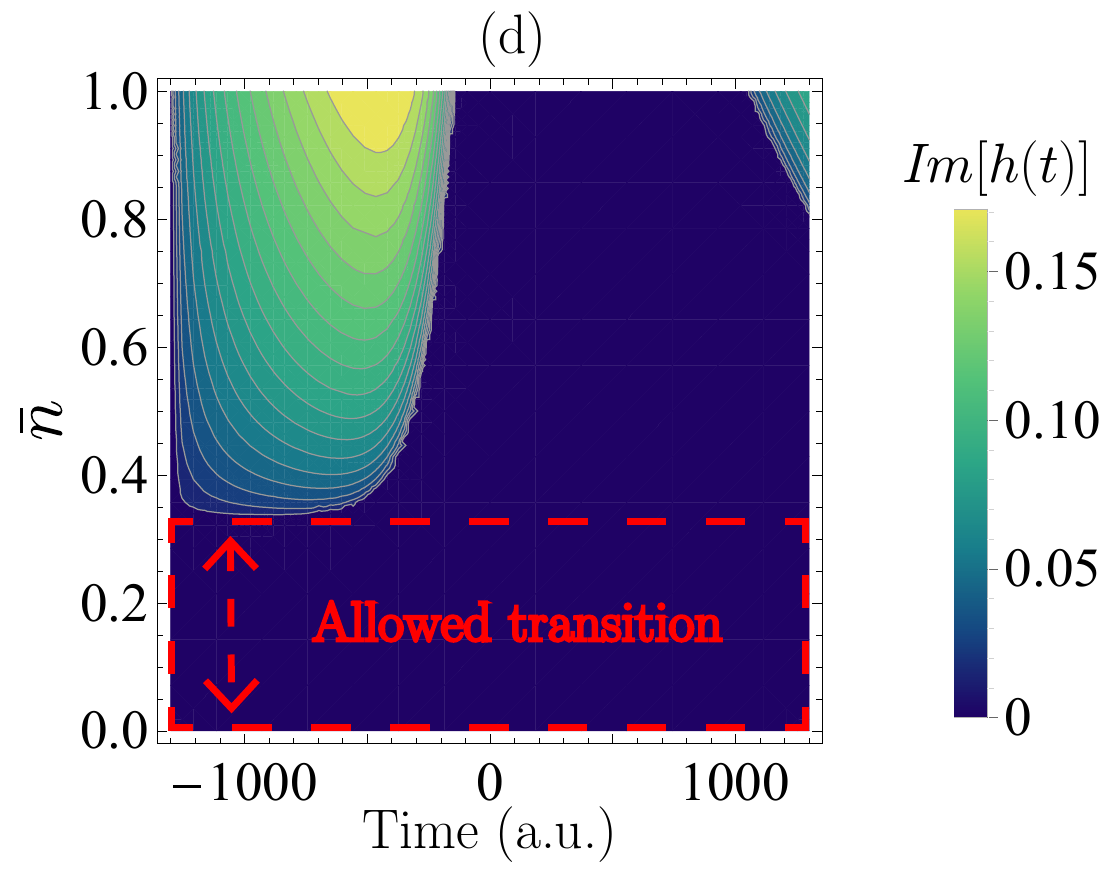}

\caption{(Color online)\textbf{ }  \textbf{Panel (a): } Imaginary
part of $h(t)$ as a function of time and $a_{f}$ for $\Gamma=10^{-4}$ a.u.,  $\gamma=0$, and $\bar{n}=0$. \textbf{Panel
(b):} Imaginary part of $h(t)$ as a function of time and $a_{f}$ for  $\Gamma=10^{-3}$ a.u., $\gamma=0$, and $\bar{n}=0$.\textbf{
Panel (c):} Imaginary part of $h(t)$ as a function of time
and $a_{f}$ for $\Gamma=10^{-4}$ a.u., $\gamma=10^{-2}$ a.u.,
and $\bar{n}=0$. \textbf{Panel (d):} Imaginary part  of $h(t)$
as a function of time and $\bar{n}$. Here we kept $a_{f}=0.4$. The frames in red (dashed) show the accessible regions. For all panels $a_{i}=0.8$. \label{fig:6}}
\end{figure}
\begin{figure}
\includegraphics[scale=0.4]{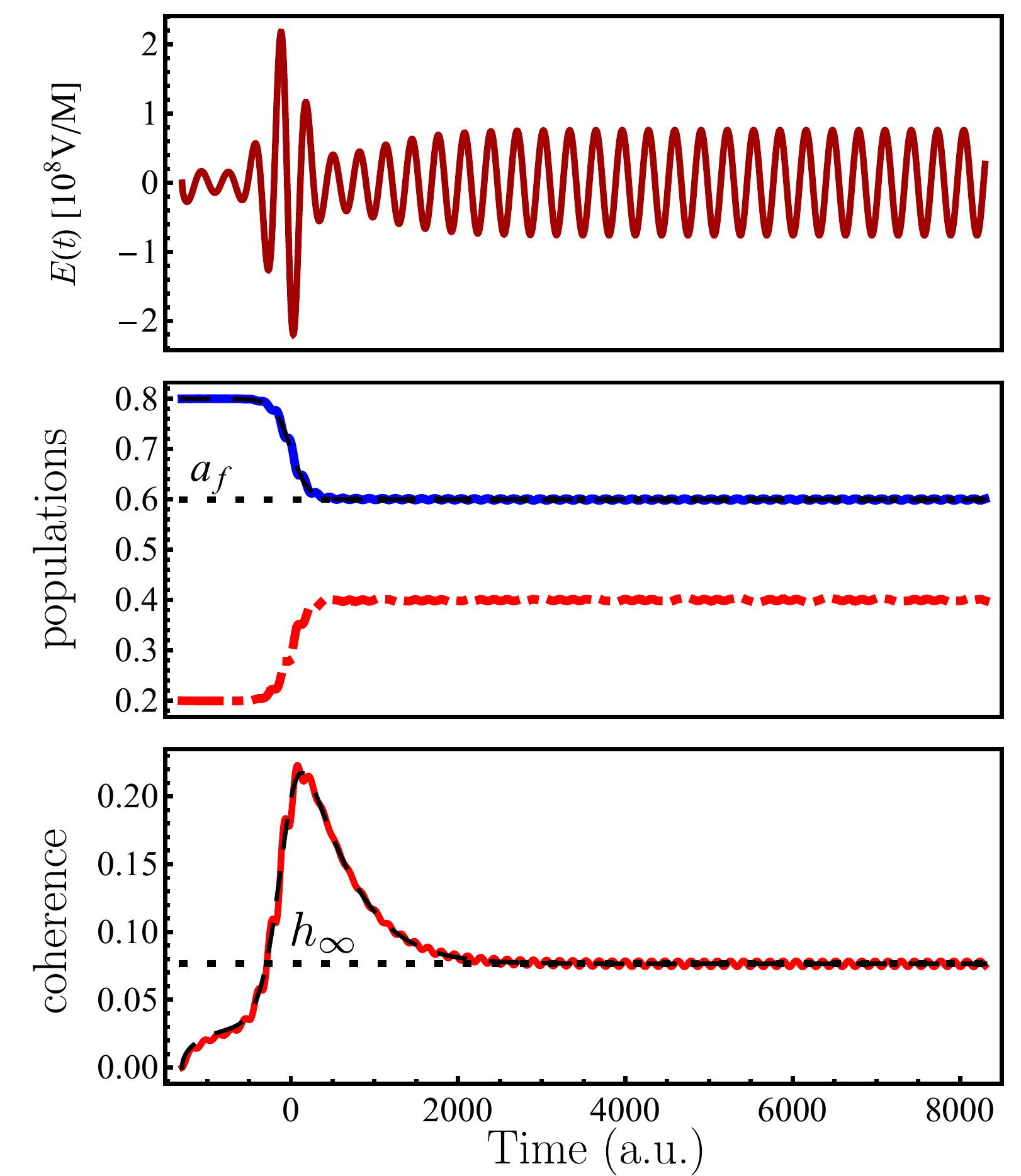}

\caption{(Color online) \textbf{Top panel}: Laser pulse as Eq.~(\ref{eq:fieldf}) (\ref{eq:fieldf}).
\textbf{Middle panel}: 
Ground state population $\rho_{gg}(t)$ is shown in blue, excited
state population $\rho_{ee}(t)$ is shown in red (dot-dashed) and
the control $f(t)$ is shown in black (dashed). \textbf{Lower panel:
}The absolute value of the coherence $|\rho_{eg}(t)|=|\rho_{ge}(t)|$ are shown in red and the
function $h(t)$ is shown in black (dashed). The dotted horizontal
line shows the steady state value of coherence $h_{\infty}$ obtained from Eq.~(\ref{eq:statcohe}). For all plots, $a_{i}=0.8$,
$a_{f}=0.6$, $\Gamma=10^{-4}$ a.u., $\gamma=10^{-3}$ a.u., $\bar{n}=0.3$.
 \label{fig:7}}
\end{figure}

In Fig. \ref{fig:6} we deepen our discussion about the achievable final populations
by showing contour plots of $h(t)$ either as a function
of time and the target population in panels (a), (b), and (c) or as a function of time and average number of thermal phonons.  We fixed the initial population as $a_{i}=0.8$
for all plots. In the panels (a) and (b) we have  dissipation rates $\Gamma=10^{-4}$ a.u. and  $\Gamma=10^{-3}$ a.u., respectively, while keeping $\gamma=0$ and $\bar{n}=0$.
We can see that by increasing $\Gamma$ we substantially change the region
of accessible populations. Interestingly, the region is reduced as well as displaced to the ground state ($a_{f}=1$). This happens
because when we increase $\Gamma$, the dissipative part of the dynamics
tries harder to push the system to the ground state (in the case $\bar{n}=0$), preventing the access to the excited state. In panel (c), we fix $\Gamma=10^{-4}$ a.u.
but add now some dephasing $\gamma=10^{-2}$ a.u., while still keeping $\bar{n}=0$. Compared to panel (a), we
see that the presence dephasing reduces the range of accessible final populations as expected given our previous analysis of Fig. \ref{fig:2}. In the last panel (d), we finally spot the role played by the temperature. The practical consequences of this analysis are far-reaching given that one usually deals with samples which are not at $T=0$ (or equivalently $\bar{n}=0$). We now fix $\Gamma=10^{-4}$ a.u.,
$\gamma=0$ and kept $a_{f}=0.4$ fixed. For the case shown in panel (d), we notice that there is an upper bound for the temperature above which it is impossible to realize the protocol. This bound is $\bar{n}\approx0.35$. Other choices of $a_i$ and parameters $\Gamma$ and $\gamma$ will lead to other upper bounds in $\bar{n}$.

\textit{Existence of steady states} -  Until now, our examples showed instances where the electromagnetic field was unable to keep the coherences. Interestingly enough, there are regimes of operation where an steady state can be sustained by the external electromagnetic field given Eq.~(\ref{eq:fieldf}). Moreover, the field can sustain the system coherence indefinitely even in the presence of dephasing and thermal noise.  By assuming the existence of such steady states, Eq.~(\ref{eq:difh}) gives us
\begin{equation}
h_{\infty}=\sqrt{\frac{(2a_{f}-1)[1+\bar{n}-(2\bar{n}+1)a_f]}{\gamma+(2\bar{n}+1)\Gamma}\Gamma},\label{eq:statcohe}
\end{equation}
where $h_{\infty}=h(t\rightarrow\infty)$. For a two-level system, $0\leq h_{\infty}\leq1/2$, and this fixes the physics  in Eq.(\ref{eq:difh}). In other words, the steady state will be reached only if
\begin{equation}\label{eq:ineq}
0\leq [2a_f-1][1+\bar{n}-(2\bar{n}+1)a_f] \leq \frac{\gamma+(2\bar{n}+1)\Gamma}{4\Gamma}.
\end{equation}
As an example, let us consider $\Gamma=10^{-4}$ a.u., $\gamma=10^{-3}$ a.u. and $\bar{n}=0.3$. Inequality (\ref{eq:ineq}) leads to $1/2\leq a_f\leq 13/16$ as achievable final populations. For $a_{i}=0.8$ and $a_{f}=0.6$, the results are shown in Fig.~\ref{fig:7}, where one can clearly see the establishment of a steady state. According to Eq. (\ref{eq:fieldf}), the field for long times oscillates with constant amplitude $2\Gamma[\mu h_{\infty}]^{-1}[1+\bar{n}-(2\bar{n}+1)a_{f}]$.
Populations and coherence also correctly follow the path given by $f(t)$ and
$h(t)$, respectively. The latter approaches the stationary value $h_{\infty}$ as predicted in Eq.~(\ref{eq:statcohe}). This protocol might find applications in the preservation of coherence in quantum memories for quantum computing \cite{U,us}.

In conclusion, we have used reverse-engineering to build a pulse able to control the populations of a two-level system
interacting with an environment. We establish a direct connection
between impossible protocols and imaginary parts of the quantum coherence.  Quite interesting, we showed how one can design a steady state protocol where coherence can be maintained indefinitely. Finally, we would like to make a few comments on the validity of the RWA in the presence of dephasing and dissipation. According to Eq. (\ref{eq:fieldf}), the stronger the environment couples to the system, the more intense the field has to be. However, for very intense fields,
the RWA tends to fail \cite{BE}. For the specific example of charge
migration in the MePeNNA molecule we  considered here, we found that the RWA works very well for dephasing rates $\gamma$ up to $10^{-2}$ a.u.
and dissipation rates $\Gamma$ up to $10^{-3}$ a.u.. This is quite a broad range if we remember that for this molecule $\omega$ is about $10^{-2}$ a.u..

\textit{Acknowledgements} - I. M. acknowledges support by the Coordenação de Aperfeiçoamento de Pessoal de Nível Superior (CAPES) and F. L. S
acknowledges partial support from of the Brazilian National
Institute of Science and Technology of Quantum Information
(INCT-IQ) and CNPq (Grant No. 302900/2017-9).

\end{document}